\documentclass[a4paper,11pt]{article}
\usepackage{graphicx}
\usepackage{amsmath}
\usepackage{times}
\usepackage{amsthm}
\usepackage{amstext}
\usepackage{amsopn}
\usepackage{amssymb}
\usepackage[section]{algorithm}
\usepackage[noend]{algpseudocode}
\usepackage{eucal}
\usepackage[latin2]{inputenc}
\usepackage{latexsym}
\usepackage[dvips,a4paper,margin=1in]{geometry}
\input epsf

\title{Exponential-Time Approximation of Hard Problems}

\date{}

\author{Marek Cygan, Łukasz Kowalik, Marcin Pilipczuk and Mateusz Wykurz\thanks{
Institute of Informatics, University of Warsaw, Poland.
This research is partially supported by a grant from the Polish Ministry of Science and Higher
Education, project N206 005 32/0807.
E-mail addresses:
\texttt{cygan@mimuw.edu.pl}, \texttt{kowalik@mimuw.edu.pl}, \texttt{malcin@mimuw.edu.pl}, \texttt{mateusz.wykurz@students.mimuw.edu.pl}.
}}

\newcommand{\ignore}[1]{}

\newcommand{\heading}[1]{
  \medskip
  \noindent \textbf{#1}
}

\floatname{algorithm}{Pseudocode}

\newtheorem {theorem}{Theorem}[section]
\newtheorem {proposition}[theorem]{Proposition}
\newtheorem {lemma}[theorem]{Lemma}
\newtheorem {corollary}[theorem]{Corollary}

\theoremstyle {definition}

\begin{document}

\maketitle

\begin{abstract}
We study optimization problems that are neither approximable in polynomial time (at least with a constant factor) nor fixed parameter tractable, under widely believed complexity assumptions.
Specifically, we focus on
 {\sc Maximum Independent Set}, {\sc Vertex Coloring}, {\sc Set Cover}, and {\sc Bandwidth}.

In recent years, many researchers design exact exponential-time algorithms for these and other hard problems. The goal is getting the time complexity still of order $O(c^n)$, but with the constant $c$ as small as possible. 
In this work we extend this line of research and we investigate whether the constant $c$ can be made even smaller when one allows constant factor approximation.
In fact, we describe a kind of approximation schemes --- trade-offs between approximation factor and the time complexity.

We study two natural approaches.
The first approach consists of designing a backtracking algorithm with a small search tree. 
We present one result of that kind: a $(4r-1)$-approximation of {\sc Bandwidth} in time $O^*(2^{n/r})$, for any positive integer $r$.

The second approach uses general transformations from exponential-time exact algorithms to approximations that are faster but still exponential-time. 
For example, we show that for any reduction rate $r$, one can transform any $O^*(c^n)$-time\footnote{$O^*(f(n))$ notation suppresses polynomial factors} algorithm for {\sc Set Cover} into a $(1+\ln r)$-approximation algorithm running in time $O^*(c^{n/r})$.
We believe that results of that kind extend the applicability of exact algorithms for NP-hard problems.

\end{abstract}

\vspace{2mm}
\noindent
{\bf Classification: Algorithms and data structures; ``fast'' exponential-time algorithms}

\section{Introduction}

\heading{Motivation}
%Many natural optimization problems are known to be NP-hard. 
%After decades of failed attempts to solve any NP-hard problem in polynomial time it seems highly %unlikely for such algorithms to exist. 
One way of coping with NP-hardness is polynomial-time approximation,\ i.e. looking for solutions that are relatively close to optimal. 
Unfortunately it turns out that there are still many problems which do not allow for good approximation. 
Let us recall some examples.
H\aa{}stad~\cite{hastad:mis} showed that {\sc Independent Set} cannot be approximated in polynomial time with factor $n^{1-\epsilon}$ for any $\epsilon>0$ unless ${\rm NP}={\rm ZPP}$. 
The same holds for {\sc Vertex Coloring} due to Feige and Kilian~\cite{feige:color-inapprox}.
By another result of Feige~\cite{feige:setcover}, {\sc Set Cover} cannot be approximated in polynomial time with factor $(1-\epsilon)\ln n$, where $n$ is the size of the set to cover, for any $\epsilon>0$ unless ${\rm NP}\subseteq{\rm DTIME}(n^{\log\log n})$. 

Another approach is the area of parametrized complexity (see e.g.~\cite{downey-fellows}). 
Then the goal is to find an algorithm with time exponential only in a parameter unrelated to the instance size (then we say the problem is {\em fixed parameter tractable}, FPT in short).
This parameter may reflect complexity of the instance -- like treewidth, but then we get an efficient algorithm only for some subclass of possible instances.
Another choice of the parameter is the measure of the solution quality. 
For example, one can verify whether in an $n$-vertex graph there is a vertex cover of size $k$ in $O(1.2738^k + kn)$ time~\cite{Chen:vtxcover}.
%Algorithms of that kind for maximization problems are even more useful since they allow for finding the solution of the given quality $k$ if one exists -- e.g.\ one can find a path of length $k$ in an $n$-vertex graph in $2^{O(k)}n^{O(1)}$ time~\cite{ayz:color-coding}.
Again, the parametrized approach does not succeed in some cases. Verifying whether a graph is $k$-colorable is NP-complete for any $k\ge 3$, while {\sc Independent Set} and {\sc Set Cover} are $W[1]$- and $W[2]$-complete respectively, meaning roughly that an FPT algorithm for {\sc Independent Set} or {\sc Set Cover} would imply algorithms of that kind for a host of other hard problems.

The aforementioned hardness results motivate the study of ``moderately exponential time'' algorithms. 
The goal here is to devise algorithms with exponential running time $O(2^{n/r})$ with $r$ big enough.
Indeed, a $O(2^{n/50})$-time algorithm may appear practical for some range of $n$, say $n\le 1000$. 
Despite some progress in this area we are still far from exact algorithms with time complexity of that order. 
One of the most researched problems in this field is {\sc Independent Set}. 
Exhaustive search for that problem gives $O(2^n)$ time bound while the currently best published result~\cite{fgk:mis} is $O(2^{n/{3.47}})$.
For {\sc Vertex Coloring} the first $O^*(2^{n/{0.77}})$-time algorithm by Lawler was then improved in a series of papers culminating in a breakthrough $O^*(2^n)$ bound of Bj\"orklund, Husfeldt and Koivisto~\cite{bhk:color-j}.

Now consider the {\sc Unweighted Set Cover} problem. 
The instance consists of a family of sets $\mathcal{S}=\{S_1,\ldots,S_m\}$.
The set $U=\bigcup\mathcal{S}$ is called \emph{the universe} and we denote $n=|U|$.
The goal is to find the smallest possible subfamily $\mathcal{C}\subseteq\mathcal{S}$ such that $\bigcup {\cal C} = U$.
Assume that $m$ is relatively small but big enough that finding an optimal solution using an exact algorithm is out of question, say $m=150$.
If the universe is small we can get a good approximation by the greedy algorithm (see e.g.~\cite{vazirani})
%~\cite{chvatal-greedy,johnson-greedy,lovasz-greedy}
with approximation ratio $H_{n} < \ln n + 1$.
However this approximation guarantee becomes bad when $n$ is big.
A natural thing to consider is an approximation algorithm with better (e.g.\ constant) guarantee and with running time exponential but substantially lower than the best known exact algorithm. 
In this paper we explore such approach. Ideally, one would like to have a kind of trade-off between the running-time and approximation ratio -- then one gets as much accuracy as can be afforded.
We study two approaches yielding results of that kind.

%\subsection{Search Tree Techniques}
\heading{Search Tree Techniques}
Many exponential-time algorithms (e.g.\ backtracking algorithms) can be viewed as visiting the nodes of an exponential sized search tree. The nodes of the tree correspond to instances of the problem, and typically the instances in the leaves are either trivial or at least solvable in polynomial-time. 

A natural idea is to use a search tree which has fewer nodes than the search tree of the exact algorithm, and with leaves corresponding to instances that can be {\em approximated} in polynomial time. This natural approach was used in a work on {\sc MAX SAT} by Dantsin, Gavrilovich, Hirsch and Konev~\cite{dantsin}.

In this paper we describe one result of that kind: a $(4r-1)$-approximation of {\sc Bandwidth} in time $O^*(2^{n/r})$, for any positive integer $r$ (see Section~\ref{sec:bandwidth} for the definition of the problem and a brief discussion of known results).

This approach can be used also for {\sc Independent Set} and {\sc Set Cover}. 
For example, {\sc Independent Set} has constant ratio approximation for graphs of bounded degree.
A standard approach to exact algorithm for this problem (used e.g.\ in~\cite{fgk:mis}) is a recursive algorithm, which picks a vertex $v$ and checks two possibilities: (1) $v$ is in the independent set -- then removes $v$ and its neighbors and makes a recursive call, and (2) $v$ is not in the independent set -- then removes $v$ and makes a recursive call.
When we can assume that the vertex picked is of large degree (if there is no such vertex polynomial-time approximation is used) we always get a big reduction of the instance size in one of the recursive calls which results in a better time bound than the exact algorithm.
However, we will not elaborate on this because we obtained much better results (also for {\sc Set Cover}) using the approach of reduction (see the next paragraph).

%\subsection{Reducibility}
\heading{Reductions}
Consider an instance $\mathcal{S}=\{S_1,\ldots,S_m\}$ of the {\sc Unweighted Set Cover} problem
described above. 
Assume $m$ is even.
Then create a new instance $\mathcal{Z}=\{Z_1,\ldots,Z_{m/2}\}$ where $Z_i=S_{2i-1}\cup S_{2i}$.
Next find an optimal solution ${\rm OPT}_{\mathcal{Z}}$ for $\mathcal{Z}$ using an exact algorithm.
Let $\mathcal{C}=\{S_{2i-1}\ |\ Z_i\in {\rm OPT}_{\mathcal{Z}}\} \cup \{S_{2i}\ |\ Z_i\in {\rm OPT}_{\mathcal{Z}}\}$.
Clearly, $|{\rm OPT}_{\mathcal{Z}}|\le |{\rm OPT}_{\mathcal{S}}|$ and hence $|\mathcal{C}|\le 2|{\rm OPT}_{\mathcal{S}}|$.
Thus we get a 2-approximation in $T(m/2)$ time where $T(m)$ is the best known bound for an exact algorithm, so currently just $T(m/2)=O^*(2^{m/2})$ after applying the exhaustive search. 
Of course this method is scalable -- similarly we get a 5-approximation in time $O^*(2^{m/5})$. 
With the present computing speed it should allow to process instances with roughly 150 sets, even when the universe $U$ is large.

The above phenomenon is the second of the two approaches studied this paper. 
We will call it a reduction.
Let us state a more precise definition now (however, we will make it a bit more general further in the paper).
Consider a minimization problem $P$ (reduction of a maximization problem is defined analogously) with a measure of instance size $s(I)$, and a measure of the solution quality $m(S)$. 
Let ${\rm OPT}_I$ be an optimal solution for instance $I$.

Let $r,a>1$ be constants.
A $(r,a)$-\emph{reduction} (or simply a \emph{reduction}) of problem $P$ is a pair of algorithms called \emph{reducer} and \emph{merger} satisfying  the following properties:
\begin{itemize}
\item 
Reducer transforms $I$, an instance of $P$, into a set $\{I_1, ..., I_k\}$ of instances of $P$  so that for every $i=1,\ldots,k$, $s(I_i)\le s(I)/r+O(1)$.
\item
Let $S_1, \ldots, S_k$ be optimal solutions of instances $I_1, \ldots, I_k$. 
Then Merger transforms the solutions $S_i$ into $S$, a solution for instance $I$, so that $m(S) \le a\cdot {\rm OPT}_I$. (Merger may also use $I$ and any information computed by reducer).
\end{itemize}

The constants $r$ and $a$ in the above definition will be called the \emph{rate} and the \emph{approximation} of the reduction, respectively
(we assume $a\ge 1$, even for a maximization problem -- then a solution with quality 
$\ge {\rm OPT}_I/a$ is returned).
Observe that above we described a $(2,2)$-reduction of {\sc Unweighted Set Cover}.
Actually, we noted that it generalizes to $(r,r)$-reduction for any $r \in \mathbb{N}$. 
If there is a $(r,a(r))$-reduction for $r$ being arbitrarily big we talk about \emph{reduction scheme} and the function $a(r)$ is the \emph{approximation} of this scheme.
(This definition is very flexible, however, most of our schemes imply a reduction for any $r \in \mathbb{N}$, and sometimes even for any $r \in \mathbb{Q}$).

\begin{table}[ht]
\begin{center}
\begin{minipage}{\textwidth}
\begin{tabular}{|l|c|c|l|} \hline
Problem & Approximation & Range of $r$ & Current time bound \\ 
\hline \hline
{\sc Unweighted Set Cover} & $1+\ln r$ & $r\in\mathbb{Q}, r\ge 1$ & $O^*(2^{n/r})$,~\cite{bhk:color-j}\\ \hline
{\sc Set Cover} & $1+\ln r$ & $r\in\mathbb{Q}, r\ge 1$ & $O^*(2^{n/{(0.5r)}}m^{\log n})$, \\ 
& & & $O^*(2^{n/(0.31r)})$, App. \ref{sec:exact-setcover}\\ \hline
{\sc Set Cover} & $r$ & $r\in\mathbb{N}, r\ge 1$ & $O^*(2^{m/r})$,~[folklore]\\ \hline
{\sc Min Dominating Set} & $r$ & $r\in\mathbb{N}, r\ge 1$ & $O^*(2^{0.598n/r})$,~\cite{rooij:mds}\\ \hline
{\sc Max Independent Set} & $r$ & $r\in\mathbb{Q}, r\ge 1$ & $O^*(2^{n/(3.47r)})$,~\cite{fgk:mis}\\ \hline
{\sc Coloring}\footnote{This is an $O^*(2^{n/3.47})$-time reduction by Bj\"orklund and Husfeldt, see~\cite{bhk:color-j}} & $1+\ln r$ & $r\in\mathbb{Q}\cap[1,4.05]$ &  $O^*(2^{n/(0.85r)})$~\cite{bhk:color-j}\\ \hline
{\sc Coloring}\footnote{This is an $O^*(2^{n/(0.85r)})$-time reduction.} & $1+0.247r\ln r$ & $r\in\mathbb{Q}, r>4.05$  & $O^*(2^{n/(0.85r)})$~\cite{bhk:color-j}\\ \hline
{\sc Coloring} & $r$ & $r\in\mathbb{N}, r>1$  & $O^*(2^{n/(0.85r)})$~\cite{bhk:color-j}\\ \hline
{\sc Bandwidth} & $r^{\log_29}=9^k$ & $r=2^k,k\in\mathbb{N}$ & $O^*(10^{n/r})$,~\cite{feige:exp}\\ \hline
{\sc Semi-Metric TSP} & $1+\log_2r=1+k$ & $r=2^k,k\in\mathbb{N}$ & $O^*(2^{n/(0.5r)})$,~\cite{bjohus:icalp}\\ \hline
\end{tabular}
\end{minipage}
\end{center}
\caption{Our reductions. Last column shows time bounds of approximation algorithms obtained using the best known (polynomial space) exact algorithms.
}
\label{tab:results}
\end{table}

We present reduction schemes for several most natural optimization problems that are both hard to approximate and resist FPT algorithms. 
Table~\ref{tab:results} shows our results.
In last column we put time bounds of approximation algorithms obtained by using our reductions with best known exact algorithms. 
As our motivations are partially practical, all these bounds refer to \emph{polynomial space} exact algorithms (putting $r=1$ gives the time complexity of the relevant exact algorithm). 
In most cases ({\sc Bandwidth}, {\sc Coloring}, {\sc Independent Set}, {\sc Semi-Metric TSP}) there are faster exact exponential-space algorithms (hence we would get also faster approximations)\ignore{\footnote{Among them, there is a recent $O(5^n)$-time, $O^*(2^n)$-space algorithm~\cite{bandwidth-wg} by the first and the third author of the present paper. Since the space complexity of this algorithm is much lower than the time complexity, it should not be a bottleneck in implementations. After mixing it with the reduction described here, we get a $9^k$-approximation in $O(5^{n/{2^k}})$-time.  However, for this problem we get a better result via different approach.}}.

Note that by putting $r=n/\log n$ we get polynomial time approximations, and for {\sc Set Cover}
we get the approximation ratio $1+\ln n$, which roughly matches the ratio of the (essentially optimal) greedy algorithm. 
Thus, our reduction can be viewed as a continuous scaling between the best possible polynomial time approximation and the best known exponential time algorithm. 
In other words, one can get as good solution as he can afford, by using as much time as he can.
A similar phenomenon appears in the case of {\sc Semi-Metric TSP}.
(This is not very surprising since these two reductions are based on the relevant polynomial time approximations). 

The notion of reduction introduced in our paper is so natural that some reductions must have been considered before, especially for big reduction rates, like $r=n/\log n$, when the reductions essentially imply polynomial-time approximation algorithms. 
We are aware of one reduction described in the context of exponential-time approximation: Bj\"o{}rklund and Husfeldt~\cite{bhk:color-j} described a reduction scheme with approximation $a(r)=(1+\ln r)$ (worth using only for bounded values of $r$ -- see \S\ref{sec:coloring}).

%\subsection{Related Work}
\heading{Related Work}
We have already mentioned results of Bj\"o{}rklund and Husfeldt~\cite{bhk:color-j} and Dantsin et\ al.~\cite{dantsin} on exponential-time approximation.
The idea of joining the worlds of approximation algorithms and ``moderately'' exponential algorithms appeared also in a recent work of Vassilevska, Williams and Woo~\cite{hybrid}, however their direction of research is completely different from ours, i.e.\ they consider so-called hybrid algorithms.
For example, they report to have an algorithm for {\sc bandwidth} which for given input {\em either} returns an $O(\log n)$-approximate solution in polynomial time {\em or} returns a $(1+\epsilon)$-approximate solution in $O(2^{n/\log\log n})$ time. 
We see that the hybrid algorithm does not guarantee constant approximation ratio and hence cannot be directly compared with our work.

Another promising area is joining the worlds of parametrized complexity and polynomial-time approximation algorithms --- see the survey paper~\cite{marx:survey}.

%\subsection{Organization of the Paper}
\heading{Organization of the Paper}
We start from the approximation scheme for the {\sc Bandwidth} problem in Section~\ref{sec:bandwidth}.
Then in Section~\ref{sec:reduction} we introduce slightly more general definition of reduction and then in Section~\ref{sec:setcover} describe two reductions for {\sc Set Cover}.
(The reductions for {\sc Maximum Independent Set}, {\sc Vertex Coloring}, {\sc Bandwidth} and {\sc Semi-Metric TSP} are put in Appendix due to space limitations.)
We conclude in Section~\ref{sec:hardness} by some complexity remarks that show relations between the notion of reduction and polynomial-time approximation.

\section{Bandwidth}
\label{sec:bandwidth}

Let $G=(V,E)$ be an undirected graph.
For a given \emph{ordering of vertices}, i.e.\ a one-to-one function $f:V\rightarrow\{1,\ldots,n\}$, its {\em bandwidth} is the maximum difference between
the numbers assigned to the endpoints of an edge, i.e.\ $\max_{uv\in E}|f(u)-f(v)|$.
{\em Bandwidth} of graph $G$, denoted by ${\rm bw}(G)$, is the minimum possible bandwidth of an ordering.
The {\sc Bandwidth} problem asks to find, for a given graph, its bandwidth with the corresponding ordering.

{\sc Bandwidth} is a notorious NP-hard problem. 
It was shown by Unger~\cite{unger:bandwidth} that {\sc Bandwidth} does not belong to APX even in
very restricted case when $G$ is a caterpillar, i.e.\ a very simple tree.
It is also hard for any fixed level of the W hierarchy~\cite{fellows:hardness}.
The best known polynomial-time approximation, due to Feige~\cite{feige-bandwidth-approx}, has $O(\log^3n\sqrt{\log n \log\log n})$ approximation guarantee. 
The fastest known exact algorithm works in time $O^*(5^n)$ and space $O^*(2^n)$ and it is due to Cygan and Pilipczuk~\cite{bandwidth-wg}, while the best polynomial-space exact algorithm, due to Feige and Kilian~\cite{feige:exp}, has time complexity $O^*(10^n)$.

We were able to find a reduction for the {\sc Bandwidth} problem. 
Although it is probably the most nontrivial of our reductions, it gives, for any $k\in \mathbb{N}$, approximation ratio $9^k$ with reduction rate $2^k$, which is far from being practical (see Appendix~\ref{sec:app:approx-bandwidth} for the details). As a corollary it gives a $9^k$-approximation in time $O^*(10^{n/2^k})$ and polynomial space, or in time $O^*(5^{n/2^k})$ and $O^*(2^{n/2^k})$ space. It is an interesting open problem whether there is a better reduction for this problem.

Now we will describe a better approximation scheme using the approach of small search tree.

\subsection{$2$-approximation in $O^*(3^n)$-time (warm-up)}
\label{sec:2-aprox}

We begin with an algorithm which is very close to a fragment of the $O^*(10^n)$-time exact algorithm of Feige and Kilian~\cite{feige:exp}.
Assume w.l.o.g.\ the input graph is connected (we keep this assumption also in the following sections).

Let $b$ be the bandwidth of the input graph --- we may assume it is given, otherwise with just a $O(\log n)$ overhead using binary search one can find the smallest $b$ for which the algorithm returns a solution.

Let us partition the set of positions $\{1,\ldots,n\}$ into $\lceil n/b\rceil$ intervals of size $b$ (except, possibly, for the last interval), so that for $j=0,\ldots,\lceil n/b \rceil - 1$, the $j$-th interval consists of positions $I_j = \{jb+1, jb+2, \ldots, (j+1)b\} \cap \{1,\ldots,n\}$.

\begin{algorithm}
\caption{\label{alg:2-apx}Generating at most $n3^n$ assignments in the $2$-approximation algorithm.}
\begin{minipage}{\textwidth}
%{\bf procedure} {\sc GenerateAssignments}($A$)
%\small
\begin{algorithmic}[1]
\Procedure{\sc GenerateAssignments}{$A$}
\If{all nodes in are assigned}
  \State If each interval $I_j$ is assigned $|I_j|$ vertices, order the vertices in intervals arbitrarily and return the ordering.
\Else
  \State $v \gets$ a vertex with a neighbor $w$ already assigned.
  \State {\bf if} $A(w)> 0$ {\bf then} {\sc GenerateAssignments}($A\cup\{(v,A(w)-1)\}$)
  \State {\sc GenerateAssignments}($A\cup\{(v,A(w))\}$)
  \State {\bf if} $A(w)<\lceil n/b \rceil-1$ {\bf then} {\sc GenerateAssignments}($A\cup\{(v,A(w)+1)\}$)
\EndIf
\EndProcedure
\vspace{2mm}
\Procedure{\sc Main}{}
\For{$j\gets 0$ {\bf to} $\lceil n/b\rceil-1$}
  \State {\sc GenerateAssignments} ($\{(r,j)\}$) \Comment{Generate all assignments with $r$ in interval $I_j$}
\EndFor
\EndProcedure
\end{algorithmic}
\end{minipage}
\end{algorithm}

The algorithm finds a set of assignments of vertices into intervals $I_j$ such that if there is an ordering $\pi$ of bandwidth at most $b$, for at least one of these assignments, for every vertex $v$, $\pi(v)$ lies in the assigned interval.
Clearly, if there is an ordering of bandwidth $b$, at least one such assignment exists.
The following method (introduced by Feige and Kilian originally for intervals of length $b/2$)
finds the required set with only at most $n3^n$ assignments.
Choose an interval for the first vertex $r$ (chosen arbitrarily) in all $\lceil n/b\rceil$ ways. 
Then pick vertices one by one, each time taking a vertex adjacent to an already assigned one.
Then there are at most 3 intervals where the new vertex can be put and so on. 
(See Pseudocode~\ref{alg:3-apx}. (Partial) assignment is represented by a set of pairs; a pair $(v,j)$ means that $A(v)=I_j$.)

Obviously, when there are more vertices assigned to an interval than its length, the assignment is skipped. Now it is clear that for any remaining assignment {\em any} ordering of the vertices inside intervals gives an ordering of bandwidth at most $2b$.
Hence we have a $3$-approximation in $O^*(3^n)$-time and polynomial space. 

Note also that if we use intervals of length $b/2$ as in Feige and Kilian's algorithm, similar method gives $3/2$-approximation in $O^*(5^n)$-time.

\subsection{Introducing the Framework}

We are going to extend the idea from the preceding section further. 
To this end we need generalized versions of simple tools used in the $2$-approximation above.

Let a {\em (partial) interval assignment} be any (partial) function $A: V \rightarrow 2^{\{1,\ldots,n\}}$, that assigns intervals of positions to vertices of the input graph. 
An interval $\{i, i+1, \ldots, j\}$ will be denoted $[i,j]$.
The {\em size} of an interval is simply the number of its elements.
Let $\pi: V \rightarrow\{1,\ldots,n\}$ be an ordering. 
When for every vertex $v$, $\pi(v)\in A(v)$, we will say that an interval assignment $A$ is {\em consistent} with $\pi$ and $\pi$ is consistent with $A$.

In Section~\ref{sec:2-aprox} all intervals had the same size $b$, moreover two intervals were always either equal or disjoint. When this latter condition holds, it is trivial to verify whether there is an assignment consistent a given interval assignment --- it suffices to check whether the number of vertices assigned to any interval does not exceed its size. 
Luckily, in a general case it is still possible in polynomial time: just note that this is a special case of scheduling jobs on a single machine with release and deadline times specified for each job (see e.g.~\cite{kleinberg}, Sec.\ 4.2) --- hence we can use the simple greedy algorithm which processes vertices in the order of $\max A(v)$ and assigns each vertex to the smallest available position.

\begin{proposition}
\label{prop:scheduling}
For any interval assignment $A$ one can verify in $O(n\log n)$ time whether there is an ordering consistent with $A$.
\end{proposition}

\ignore{
\begin{proof}
%The ordering corresponds to a perfect matching in the bipartite graph $B=(V_1,V_2,E_B)$, with %$V_1=V$, $V_2=\{1,\ldots,n\}$ such that $vk \in E_B$ iff $k \in A(v)$.
%Such a matching can be found in polynomial time by the classical Hopcroft-Karp algorithm.
%However, we note here that in the case 
We use the following simple, greedy algorithm, that assigns vertices to successive positions $1,2, \ldots, n$.  For any position $i$, it chooses the vertex $v$ such that $i\in A(v)$ and for any other vertex $w$ with $i\in A(w)$, $\max A(v) \le \max A(w)$.
\end{proof}
}

To get a nice approximation ratio, however, we need a bound on the bandwidth of resulting ordering.
The obvious bound is \[\max_{uv\in E} \max_{i\in A(u), \atop j\in A(v)} |i-j|.\]
The above bound was sufficient in Section~\ref{sec:2-aprox}, but we will need a better bound.

\begin{lemma}
\label{lem:s+b}
Let $A$ be an interval assignment for an input graph $G=(V,E)$.
Let $s$ be the size of the largest interval in $A$, i.e.\ $s=\max_{v\in V}|A(v)|$.
If there is an ordering $\pi^*$ of bandwidth $b$ consistent with $A$, then
one can find in polynomial time an ordering $\pi$ that is consistent with $A$ and has bandwidth at most $s+b$.
\end{lemma}

\begin{proof}
Consider any edge $uv$. 
Clearly, $\pi^*(u) \in [\min A(v) - b, \max A(v)+b]$. 
Hence we can replace  $A(u)$ by $A(u) \cap [\min A(v) - b, \max A(v)+b]$, maintaining the invariant that $\pi^*$ is consistent with $A$.
Similarly, we can replace  $A(v)$ by $A(v) \cap [\min A(u) - b, \max A(u)+b]$.
Our algorithm performs such replacements for every edge $uv\in E$.
%Our algorithm performs such replacements as long as they modify $A$.
As a result we get an assignment $A'$ such that for every edge $uv$, 
$\max_{i\in A'(u), j\in A'(v)} |i-j| \le s+b$.
It is clear that any ordering consistent with $A'$ has bandwidth at most $s+b$.
Such the ordering can be found in polynomial time by Proposition~\ref{prop:scheduling}
\end{proof}

In the following sections it will be convenient to formalize a little the order in which intervals are assigned to the vertices of the input graph.
Recall that each time an interval is assigned to a new vertex $v$, $v$ has an already assigned neighbor $w$, except for the initial vertex $r$. In other words, the algorithm builds a rooted spanning tree for each assignment (here, $r$ is the root, and $w$ is the parent of $v$).
In what follows, we will fix a rooted spanning tree $T$, and our algorithm will generate interval assignments in such a way that the first vertex assigned $r$ is the root of $T$, and whenever a vertex $v\ne r$ is assigned, its parent in $T$ has been already assigned.

\subsection{$3$-approximation in $O^*(2^n)$-time}

This time, the algorithm uses $\lceil n/b \rceil$ intervals, each of size $2b$ (except for one or two last intervals), so that for $j=0,\ldots,\lceil n/b \rceil - 1$, the $j$-th interval consists of positions $I_j = \{jb+1, jb+2, \ldots, (j+2)b\} \cap \{1,\ldots,n\}$. Note that the intervals overlap.

\begin{algorithm}
\caption{\label{alg:3-apx}Generating at most $n2^n$ assignments in the $3$-approximation algorithm.}
\begin{minipage}{\textwidth}
%{\bf procedure} {\sc GenerateAssignments}($A$)
%\small
\begin{algorithmic}[1]
\Procedure{\sc GenerateAssignments}{$A$}
\If{all nodes in $T$ are assigned}
  \State Using Lemma~\ref{lem:s+b} find ordering consistent with interval assignment corresponding to $A$.
\Else
  \State $v \gets$ a node in $T$ such that $v$'s parent $w$ is assigned.
  \State {\bf if} $A(w)> 0$ {\bf then} {\sc GenerateAssignments}($A\cup\{(v,A(w)-1)\}$)
  \State {\bf if} $A(w)<\lceil n/b \rceil-1$ {\bf then} {\sc GenerateAssignments}($A\cup\{(v,A(w)+1)\}$)
\EndIf
\EndProcedure
\vspace{2mm}
\Procedure{\sc Main}{}
\For{$j\gets 0$ {\bf to} $\lceil n/b\rceil-1$}
  \State {\sc GenerateAssignments} ($\{(r,j)\}$) \Comment{Generate all assignments with root in $I_j$}
\EndFor
\EndProcedure
\end{algorithmic}
\end{minipage}
\end{algorithm}

The algorithm generates all possible assignments of vertices to intervals in such a way
that if a node in $T$ is assigned to interval $I_j$ then each of its children is assigned to interval $I_{j-1}$ or $I_{j+1}$. Clearly, there are at most $n2^n$ such assignments.
Moreover, if there is an ordering $\pi$ of bandwidth $b$, then the algorithm generates an interval assignment $A_{\pi}$ consistent with $\pi$. 
To find $A_{\pi}$, visit the nodes of $T$ in preorder, assign the root $r$ to the interval $S_{\lfloor(\pi(r)-1)/r\rfloor}$, and for each node $v$ with parent $w$ already assigned to an interval $I_j$, put $A_{\pi}(v)=I_{j+1}$ if $\pi(v) > (j+1)b$ and $A_{\pi}(v)=I_{j-1}$ otherwise.
For each generated assignment the algorithm tries to find an ordering of bandwidth at most $3b$ using Lemma~\ref{lem:s+b}. Clearly, it succeeds for at least one assignment, namely $A_{\pi}$.
The algorithm is sketched in Pseudocode~\ref{alg:3-apx}.

\subsection{$(4r-1)$-approximation in $O^*(2^{n/r})$-time}

In this section we are going to generalize the algorithm from the prior section to an approximation scheme. 
Let $r$ be a positive integer. We will describe a $(4r-1)$-approximation algorithm.
Our algorithm uses intervals of sizes $2ib$, for $r \le i \le 2r - 1$. 
Note that unlike in previous algorithms, intervals of many different sizes are used.
As before, intervals will begin in positions $jb + 1$, for $j=0,\ldots,\lceil n/b \rceil-1$.
The interval beginning in $jb + 1$ and of length $2ib$ will be denoted by $I_{j,2i}$.
For convenience, we allow intervals not completely contained in $\{1,\ldots, n\}$, but each assigned interval contains at least one position from $\{1,\ldots,n\}$.

\begin{algorithm}
\caption{\label{alg:(4r-1)-apx}Generating assignments in the $(4r-1)$-approximation algorithm.}
\begin{minipage}{\textwidth}
%\small
\begin{algorithmic}[1]
\Procedure{\sc GenerateAssignments}{$A$}
\If{all nodes in $T$ are assigned}
   \State Cut all intervals in $A$ to make them contained in $\{1,\ldots,n\}$.
   \State Using Lemma~\ref{lem:s+b} find ordering consistent with interval assignment corresponding to $A$.
\Else
  \State $v \gets$ a node in $T$ such that $v$'s parent $w$ is assigned; let $I_{j,2i} = A(w)$.
  \If{$i+1 \le 2r-1$}
    \State{\sc GenerateAssignments}($A\cup\{(v,I_{j-1,2(i+1)}\}$)
  \Else  
    \State {\bf if} $j-1+2r \ge 1$ {\bf then} {\sc GenerateAssignments}($A\cup\{(v,I_{j-1,2r}\}$)
    \State {\bf if} $j-1+2r \le \lceil n/b \rceil - 1$ {\bf then} {\sc GenerateAssignments}($A\cup\{(v,I_{j-1+2r,2r}\}$)
  \EndIf  
\EndIf
\EndProcedure
\vspace{2mm}
\Procedure{\sc Main}{$i_0$}
\For{$j\gets 0$ {\bf to} $\lceil n/b\rceil - 1$}
  \State {\sc GenerateAssignments} ($\{(r,I_{j,2i_0})\}$) \Comment{Generate all assignments with root in $I_{j,2i_0}$}
\EndFor
\EndProcedure
\end{algorithmic}
\end{minipage}
\end{algorithm}

The algorithm is sketched in Pseudocode~\ref{alg:(4r-1)-apx}. 
Let $i_0\in\{r,\ldots,2r-1\}$ be a parameter that we will determine later.
The algorithm assigns the root of $T$ to all possible intervals of size $2i_0b$ that overlap with $\{1,\ldots,n\}$ and extends each of these partial assignments recursively.

To extend a given assignment, the algorithm chooses a node $v$ of $T$ such that the parent $w$ of $v$ has been already assigned an interval, say $I_{j,2i}$. 
Consider the interval $I_{j-1,2(i+1)}$ which is obtained from $I_{j,2i}$ by ``extending'' it by $b$ positions both at left and right side. Note that in any ordering consistent with the current assignment, $v$ is put in a position from $I_{j-1,2(i+1)}$.
Hence, if $I_{j-1,2(i+1)}$ is not too big, i.e.\ $i+1\le 2r-1$, the algorithm simply assigns $I_{j-1,2(i+1)}$ to $v$ and proceeds with no branching (just one recursive call).
Otherwise, if $i+1 = 2r$, the interval $I_{j-1,2(i+1)}$ is split into two intervals of size $2r$, namely $I_{j-1,2r}$ and $I_{j-1+2r,2r}$ and two recursive calls follow: with $v$ assigned to $I_{j-1,2r}$ and $I_{j-1+2r,2r}$ respectively.

As before, for every generated assignment (after cutting the intervals to make them contained in $\{1,\ldots,n\}$) the algorithm applies Lemma~\ref{lem:s+b} to verify whether it is consistent with an ordering of bandwidth $[2(2r-1)+1]b=(4r-1)b$. Similarly as in the case $r=1$ described before, for at least one assignment such the ordering is found.

We conclude with the time complexity analysis.
Observe that the nodes at tree distance $d$ from the root are assigned intervals of size $2[(i_0+d) \bmod r + r]$.
It follows that branching appears only when $i_0+d \equiv 0 \pmod r$. 
Let $\hat{n}(i_0)$ denote the number of nodes at tree distance $d$ satisfying this condition.
It is clear that the above algorithm works in time $O^*(2^{\hat{n}(i_0)})$.
Since $\sum_{i \in \{r,\ldots,2r-1\}}\hat{n}(i)=n$, for some $i \in \{r,\ldots,2r-1\}$, $\hat{n}(i) \le n/r$. By choosing this value as $i_0$, we get the $O^*(2^{n/r})$ time bound.

\begin{theorem}
For any positive integer $r$, there is a $(4r-1)$-approximation algorithm for {\sc Bandwidth} running in $O^*(2^{n/r})$ time and polynomial space.\qed
\end{theorem}

\section{Reducibility (slightly more general)}
\label{sec:reduction}

In this section we introduce a slightly more general version of reduction and discuss some of its basic properties. 
Essentially the difference from the version in the Introduction is that sometimes between reducer and merger we want to use approximation algorithm instead of exact one.

As before let $P$ be a minimization problem, let $s(I)$ and $m(S)$ denote the measures of the instance size and the solution quality, respectively.
Let $r>1$ be a constant and $f:\mathbb{R}\rightarrow\mathbb{R}$ a function.
An $(r,f)$-\emph{reduction} (or simply a \emph{reduction}) of $P$ is a pair of algorithms called  \emph{reducer} and \emph{merger} satisfying the following properties:
\begin{itemize}
\item 
Reducer transforms $I$, an instance of $P$, into a set $\{I_1, ..., I_k\}$ of instances of $P$  so that for every $j=1,\ldots,k$, $s(I_j)\le s(I)/r + O(1)$.
\item
Let $S_1, \ldots S_k$ be solutions of instances $I_1, \ldots, I_k$. 
Let $\alpha>1$ be an approximation guarantee of these solutions, i.e. 
for $j=1,\ldots,k$, $m(S_j)\le\alpha m({\rm OPT}_{I_j})$. 
Then merger transforms the solutions $S_i$ into $S$, a solution for instance $I$, so that $m(S) \le f(\alpha) {\rm OPT}_I$. (Merger may also use $I$ and any information computed by reducer).
\end{itemize}
As before, $r$ is called the rate and $f$ is called the approximation (since we do not expect $f$ to be a constant function it should not lead to ambiguity). 
Again, if there is a $(r,f_r)$-reduction for arbitrarily big $r$ we deal with \emph{reduction scheme} with approximation $a(r,\alpha)=f_r(\alpha)$.
Note that the already described reduction for {\sc Unweighted Set Cover} has approximation $a(r,\alpha)=r\alpha$.

The time complexity of the reduction is the sum of (worst-case) time complexities of reducer and merger. 
In most cases our reductions will be polynomial time. 
However, under some restrictions exponential-time reductions may be interesting as well. 

The following lemma will be useful (an easy proof is in Appendix~\ref{sec:app:proof-complemma}):

\begin{lemma}[Reduction Composition]
If there is an $(r,f)$-reduction $R$ then for any positive $k\in\mathbb{N}$ there is a
$(r^k,f^k)$-reduction\footnote{$f^k$ is the composition: $f^k=f\circ f^{k-1}$, $f^0=id$.} $R'$ for the same problem.
Moreover, if the merger of $R$ generates a polynomial number of instances then
$R'$ has the same time complexity as $R$, up to a polynomial factor.
\end{lemma}

Note that the Reduction Composition Lemma implies that once we find a single $(r,f)$-reduction, it extends to a reduction scheme, though for quite limited choice of $r$. 
We will see more consequences of this lemma in Section~\ref{sec:hardness}.
 
\section{Set Cover}
\label{sec:setcover}
We will use the notation for {\sc Set Cover} from the Introduction.
Since here we consider the general, weighted version of the problem, now each set $S\in\mathcal{S}$ comes with its weight $w(S)$.
We will also write $w(\mathcal{C})$ for the total weight of a family of sets $\mathcal{C}$.
In the case of {\sc Set Cover} there are two natural measures for size of the instance: the size of the universe $U$ and the number of sets in the family $\mathcal{S}$. 
We will present reductions for both measures. 
Both reductions work for the weighted version of the problem.

\subsection{Reducing the size of universe}
\label{sec:greedy}

An $r$-approximate solution of {\sc Set Cover} can be found by dividing $U$ into $r$ parts,
covering each of them separately and returning the union of these covers.
It corresponds to a reduction scheme with approximation $a(r,\alpha)=r\alpha$ for $r\in\mathbb{N}$.
However, we will describe a much better reduction scheme.

Let's recall the greedy algorithm (see e.g.~\cite{vazirani}),
%~\cite{chvatal-greedy,johnson-greedy,lovasz-greedy}
called {\sf Greedy} from now.
It selects sets to the cover one by one.
Let $\mathcal{C}$ be the family of sets chosen so far.
Then {\sf Greedy} takes a set that covers \emph{new} elements as cheap as possible, i.e.
chooses $S$ so as to minimize $w(S)/|S\setminus \bigcup \mathcal{C}|$.
For each element $e\in S\setminus \bigcup \mathcal{C}$ the amount $w(S)/|S\setminus \bigcup \mathcal{C}|$ is called the \emph{price} of $e$ and denoted as ${\rm price}(e)$. 
This procedure continues until $\mathcal{C}$ covers the whole $U$.
Let $e_1,\ldots,e_n$ be the sequence of all elements of $U$ in the order of covering by
{\sf Greedy} (ties broken arbitrarily).
The standard analysis of {\sf Greedy} uses the following lemma (see~\cite{vazirani} for the proof).

\begin{lemma}
\label{lem:greedy}
For each $k\in 1,\ldots,n$, ${\rm price}(e_k) \le w({\rm OPT})/(n-k+1)$
\end{lemma}

%\begin{proof}
%(We put the proof for completeness.)
%Consider the iteration when the element $e_k$ is covered by some set $S$.
%Before $S$ was chosen there were at least $n-k+1$ not covered elements.
%Hence $w(S)/|S\setminus \bigcup \mathcal{C}|\ge w({\rm OPT})/(n-k+1)$ since
%$\rm{OPT}$ covers these elements with average element price $w({\rm OPT})/(n-k+1)$.
%
%\end{proof}

The idea of our reduction is very simple. 
For example, assume we aim at a reduction rate 2, and $n$ is even.
Lemma~\ref{lem:greedy} tells us that {\sf Greedy} starts from covering elements very cheaply, and than pays more and more. 
So we just stop it before it pays much but after it covers sufficiently many elements.
Note that if we manage to stop it just after $e_{n/2}$ is covered the total price of the covered elements (and hence the weight of the sets chosen) is at most $(H_n-H_{n/2})w({\rm OPT})=(\ln 2 + O(1/n))w({\rm OPT})$. 
If we cover the remaining elements, say, by exact algorithm we get roughly a $(1+\ln 2)$-approximation.
However the set that covers $e_{n/2}$ may cover many elements $e_i$, $i>n/2$. 
By Lemma~\ref{lem:greedy} the price of each of them is at most $w({\rm OPT})/(n/2) = 2w({\rm OPT})/n$. 
Hence this last set costs us at most $w({\rm OPT})$ and together we get roughly a 
$(2+\ln 2)$-approximation. 
Luckily, it turns out that paying $w({\rm OPT})$ for the last set chosen by {\sf Greedy} is not necessary: below we show a refined algorithm which would yield a $(1+\ln 2)$-approximation in this particular case.

\begin{theorem}
There is a polynomial-time $|U|$-scaling reduction scheme for {\sc Set Cover} with approximation $a(r,\alpha)=\alpha+\ln r + O\left(1/n\right)$, for any $r \in \mathbb{Q}$, $r>1$. 
\end{theorem}

\begin{algorithm}
\caption{\label{alg:set-cover}Universe-scaling reducer for {\sc Set Cover}}
\begin{minipage}{\textwidth}
%\small
\begin{algorithmic}[1]
\State $\mathcal{C}\gets\emptyset$.
\While{$\bigcup\mathcal{S}\cup\bigcup\mathcal{C} = U$}
  \State Find $T\in\mathcal{S}$ so as to minimize $\frac{w(T)}{|T\setminus \bigcup\mathcal{C}|}$
  \If{$n-|\bigcup\mathcal{C}\cup T| > n/r$}
    \State $\mathcal{C}\gets\mathcal{C}\cup\{T\}$.
  \Else
    \State $\mathcal{C}_T\gets \mathcal{C}$
    \State [Create an instance $I_T=(\mathcal{S}_T,w)$:]
    \State for each $P\in\mathcal{S}$, $\mathcal{S}_T$ contains set $P\setminus(\bigcup\mathcal{C}_T\cup T)$, of weight $w(P)$.
    \State $\mathcal{S}\gets\mathcal{S}\setminus \{T\}$ \label{ln:set-cover:remove}
  \EndIf
\EndWhile
\end{algorithmic}
\end{minipage}
\end{algorithm}

\begin{proof}
Let $I=(\mathcal{S},w)$ be an instance of {\sc Set Cover} problem.
Reducer works similarly as {\sf Greedy}. 
However, before adding a set $T$ to the partial cover $\mathcal{C}$ it checks whether
adding $T$ to $\mathcal{C}$ makes the number of non-covered elements at most $n/r$.
If so, $T$ is called a {\em crossing set}. 
Instead of adding $T$ to $\mathcal{C}$, reducer creates an instance $I_T=(\mathcal{S}_T,w)$ of {\sc Set Cover} that will be used to cover the elements covered neither by $\mathcal{C}$ nor by $T$.
Namely, for each $P\in\mathcal{S}$, $\mathcal{S}_T$ contains set $P\setminus(\bigcup\mathcal{C}\cup T)$, of weight $w(P)$.
Apart from $I_T$ reducer stores $\mathcal{C}_T$, a copy of $\mathcal{C}$, which will be used by merger.
After creating the instance, set $T$ is removed from the family of available sets $\mathcal{S}$. 
If it turns out that the universe cannot be covered after removing $T$, i.e.\  $\bigcup\mathcal{S}\cup\bigcup\mathcal{C}\ne U$, the reducer stops.
See Pseudocode~\ref{alg:set-cover} for details. 
Note that reducer creates at least 1 and at most $|\mathcal{S}|$ instances.

Let $I_T$ be any instance created for some crossing set $T$ and 
let ${\rm SOL}_T\subset\mathcal{S}$ be its solution such that 
$w({\rm SOL}_T)\le\alpha w({\rm OPT}_{I_T})$, $\alpha \ge 1$.
Let $\mathcal{S}_T' = \mathcal{C}_T \cup \{T\} \cup {\rm SOL}_T$.
Clearly, $\mathcal{S}_T'$ is a cover of $U$ for every crossing set $T$.
The merger simply selects the lightest of these covers.

Let $T^*$ be the first crossing set found by reducer such that $T^*$ belongs to ${\rm OPT}_I$, some optimal solution for instance $I$ (note that at least one crossing set is in ${\rm OPT}_I$).
Clearly ${\rm OPT}_I\setminus\{T^*\}$ covers $\bigcup\mathcal{S}_{T^*}$.
Hence $w({\rm OPT}_{I_{T^*}}) \le w({\rm OPT}_I \setminus \{T^*\})$ so $w(T^*) + w({\rm OPT}_{I_{T^*}}) \le w({\rm OPT}_I)$.
It follows that $w(T^*) + w({\rm SOL}_{T^*}) \le \alpha w({\rm OPT}_I)$.
Since $\mathcal{C}_{T^*}$ covers less than $n- n/r$ elements, by Lemma~\ref{lem:greedy} 
\begin{eqnarray*}
w(\mathcal{C}_{T^*}) & \le &
\sum_{k=1}^{\lfloor n- n/r \rfloor}\frac{w({\rm OPT}_I)}{n-k+1}=
\sum_{k=1}^{n- \lceil n/r \rceil}\frac{w({\rm OPT}_I)}{n-k+1}=
(H_n-H_{\lceil n/r\rceil})w({\rm OPT}_I)=\\
&=&(\ln n - \ln\left\lceil n/r\right\rceil + O\left(1/n\right))w({\rm OPT}_I)\le 
(\ln r + O\left(1/n\right))w({\rm OPT}_I).
\end{eqnarray*}
We conclude that merger returns a cover of weight $\le(\alpha+\ln r + O\left(\tfrac{1}{n}\right)) w({\rm OPT}_I)$.
\end{proof}

Clearly, to make use of universe-scaling reduction we need a $O^*(c^n)$ exact algorithm, where $c$ is a constant. As far as we know there is no such result published. 
However we can follow the divide-and-conquer approach of Gurevich and Shelah~\cite{gurevich} rediscovered recently Bj\"orklund and Husfeldt~\cite{bjohus:icalp} and we get a $O^*(4^nm^{\log n})$-time algorithm. If $m$ is big we can use another, $O^*(9^n)$-time version of it.
See Appendix~\ref{sec:exact-setcover} for details.

We also note that for the unweighted case there is an $O(2^nmn)$-time polynomial space algorithm by Bj\"orklund et al.~\cite{bhk:color-j} using the inclusion-exclusion principle.

\subsection{Reducing the number of sets}
\label{sec:set-merge}

Recall the reduction described in the introduction. 
In the weighted version it fails, basically because the sets from the optimal solution may be joined with some heavy sets. 
The natural thing to do is sorting the sets according to their weight and joining only neighboring sets.
This simple modification does not succeed fully but with some more effort we can make it work.

\begin{theorem}
There is a polynomial-time $|\mathcal{S}|$-scaling reduction scheme for {\sc Set Cover} with approximation $a(r,\alpha)=\alpha r$, for any $r \in \mathbb{N}$, $r>1$.
\end{theorem}

\newcommand{\calS}{\mathcal{S}}
\newcommand{\calB}{\mathcal{B}}
\newcommand{\calC}{\mathcal{C}}
\newcommand{\calR}{\mathcal{R}}
\newcommand{\calX}{\mathcal{X}}

\begin{proof}
Reducer starts from sorting the sets in $\calS$ in the order of non-decreasing weight.
So let $\calS = \{S_1,\ldots,S_m\}$ so that $w(S_1)\le w(S_2) \le \ldots \le w(S_m)$.
Next it partitions this sequence into blocks $\mathcal{B}_i$, $i=1,\ldots,\lceil m/r \rceil$, 
each of size at most $r$, namely $\mathcal{B}_i=\{S_j\in\calS\ |\ (i-1)r<j\le ir\}$.
Let $U_i=\bigcup\mathcal{B}_i$ be the union of all sets in $\mathcal{B}_i$ and define its weight as the total weight of $\mathcal{B}_i$, i.e.\ $w(U_i)=w(\mathcal{B}_i)$.
For any $k=1,\ldots,m$ we also define $\calX_k=\{S_j\in \calB_{\lceil k/r\rceil} \ |\ j<k\}$
and $V_k=\bigcup \calX_k$ with $w(V_k)=w(\calX)$.
Reducer creates $m$ instances, namely 
$\calS_i=\{U_j\ |\ S_i\not\in \calB_j\}\cup\{V_i,S_i\}$ for $i=1,\ldots,m$.

Of course any subfamily (or a cover) $\mathcal{C}\subseteq\calS_i$ corresponds to 
$\widehat{\mathcal{C}}$,
%=\bigcup_{j:U_j\in\mathcal{C}}\mathcal{B}_j\cup(\mathcal{B}_i\cap\calC)$, 
a subfamily of $\calS$ with the same weight obtained from $\calC$ by splitting the previously joined sets (we will use this denotation further).
Clearly $\bigcup\calC=\bigcup\widehat{\calC}$, in particular if $\calC$ is a cover, so is $\widehat{\calC}$.

Let $\calC_1,\ldots,\calC_m$ be solutions (covers) for the instances created by reducer, such that $w(\calC_i)\le \alpha w({\rm OPT}_{\calS_i})$.
Merger simply chooses the lightest of them, say $\calC_q$, and returns $\widehat{\calC_q}$, a cover of $U$.

Now it suffices to show that one of the instances has a cover that is light enough.
Let $i^*=\max\{i\ |\ S_i\in{\rm OPT}\}$.
We focus on instance $\calS_{i^*}$.
If $\calX_{i^*}\cap {\rm OPT}=\emptyset$ we choose its cover
$\calR=\{U_j \in \calS_{i^*}\ |\ \calB_j\cap{\rm OPT} \ne \emptyset\}\cup\{S_{i^*}\}$,
otherwise
$\calR=\{U_j \in \calS_{i^*}\ |\ \calB_j\cap{\rm OPT} \ne \emptyset\}\cup\{V_{i^*},S_{i^*}\}$
Clearly it suffices to show that $w(\widehat{\calR}\setminus{\rm OPT}) \le (r-1)w({\rm OPT})$. 
Consider any $S_i \in \widehat{\calR}\setminus{\rm OPT}$.
If $S_i\not\in\calX_{i^*}$ we put 
$f(i)=\min\{j\ |\ S_j \in {\rm OPT}\text{ and }\lceil j/r \rceil>\lceil i/r \rceil\}$, otherwise $f(i)=i^*$. Then $w(S_i) \le w(S_{f(i)})$. 
We see that $f$ maps at most $r-1$ elements to a single index of a set from ${\rm OPT}$, so indeed
$w(\widehat{\calR}\setminus{\rm OPT}) \le (r-1)w({\rm OPT})$ and hence
$w(\widehat{\calR}) \le r w({\rm OPT})$.
It follows that $w({\rm OPT}_{\calS_{i^*}}) \le r w({\rm OPT})$ so 
$w(\calC_{i^*}) \le \alpha r w({\rm OPT})$ and finally $w(\widehat{\calC_{q}}) \le \alpha r w({\rm OPT})$.
\end{proof}

\subsection{Special Case: (Weighted) Minimum Dominating Set}
Of course, {\sc Minimum Dominating Set} is a special case of {\sc Set Cover} -- a graph $G=(V,E)$ corresponds to the set system $\calS=\{N[v]\ |\ v\in V\}$, where $N[v]$ consists of $v$ and its neighbors. 
Note the set merging algorithm described in Section~\ref{sec:set-merge} can be adapted here: merging sets corresponds simply to identifying vertices. Hence we get a reduction scheme with approximation $a(r,\alpha) = \alpha r$. Combined with the recent $O(2^{0.598n})$-time exact algorithm by Rooij and Bodleander we get an $r$-approximation in time $O(2^{0.598n/r})$, for any natural $r$.

\section{Reductions and polynomial-time approximation}
\label{sec:hardness}

Is it possible to improve any of the reductions presented before?
Are some of them in some sense optimal?
To address these questions at least partially we explore some connections between
reductions and polynomial-time approximation.
For example note that the $(r,\alpha r)$-reduction for {\sc Max Independent Set} implies $(n/\log n)$-approximation in polynomial time, by putting $r=n/\log n$ and using an exact algorithm for instances of size $O(\log n)$.
Since we know that {\sc Max Independent Set} cannot be approximated much better in polynomial time it suggests that this reduction may be close to optimal in some sense.
The following lemma is an immediate consequence of Reduction Composition Lemma.
Let us call a reduction {\em bounded} when the reducer creates $O(1)$ instances.

\begin{lemma}
If for some $r>1$ there is a polynomial-time bounded $(r,f)$-reduction for problem $P$, then
$P$ is $f^{log_rn - \log_r\log_2 n}(1)$-approximable in polynomial time.

\end{lemma}

\begin{corollary}
\label{cor:hardness1}
If for some constants $c,r$, $r>1,c>0$ there is a polynomial-time bounded $(r,f)$-reduction with $f(\alpha)=c\alpha+o(\alpha)$ for problem $P$, then
$P$ is $O(\frac{n}{\log n})^{\log_rc}$-approximable in polynomial time.

\end{corollary}

Note that Corollary~\ref{cor:hardness1} implies that neither {\sc Max Independent Set} nor {\sc Vertex Coloring} has a polynomial time bounded $(r,qr\alpha+o(\alpha))$-reduction for any $q<1$, unless ${\rm NP}={\rm ZPP}$. 
If we skip the assumption that the reduction is bounded, the existence of such a reduction implies approximation in $n^{O(\log n)}$ time, which is also widely believed to be unlikely.
Hence improved reductions need use either exponential time or some 
strange dependence on $\alpha$, say $a(\alpha)=0.1\alpha^2$.

\bibliographystyle{abbrv}
\bibliography{expapx}

\pagebreak

\appendix

\section{Proof of the Reduction Composition Lemma}
\label{sec:app:proof-complemma}

\begin{lemma}[Reduction Composition]
If there is a $(r,f)$-reduction $R$ then for any positive $k\in\mathbb{N}$ there is a
$(r^k,f^k)$-reduction\footnote{$f^k$ is the composition: $f^k=f\circ f^{k-1}$, $f^0=id$.} $R'$ for the same problem.
Moreover, if the merger of $R$ generates a polynomial number of instances then
$R'$ has the same time complexity as $R$, up to a polynomial factor.
\end{lemma}

\begin{proof}
We use induction on $k$. 
Let $R$ and $M$ be the reducer and the merger of $R$, respectively.
We will describe $R'$, which consists of the reducer $R_k$ and the merger $M_k$.

For $k=1$ the claim is trivial. 
Now assume there is a $(r^{k-1},f^{k-1})$-reduction $Q$ with reducer $R_{k-1}$ and merger $M_{k-1}$. 

Let $I$ be the input instance for reducer $R_k$.
$R_k$ executes $R_{k-1}$ which generates instances $I_1,\ldots,I_q$.
By induction hypothesis, $s(I_i)\le s(I)/r^{k-1}$ for $i=1,\ldots,q$.
Then for each $i=1,\ldots,q$, $R_k$ applies $R$ to $I_i$, which generates instances $I_{i,1},\ldots,I_{i,q_i}$.
Note that for each $i,j$, $s(I_{i,j})\le s(I)/r^k$.

Now assume $S_{i,j}$ is a solution for $I_{i,j}$ such that $S_{i,j} \le \alpha {\rm OPT}_{I_{i,j}}$ (for minimization problem)\footnote{The proof for a maximization problem is analogous}.
Merger $M_k$ applies $M$ to every sequence of solutions $S_{i,1},\ldots,S_{i,q_i}$ and gets a resulting solution $S_i$ for every $i=1,\ldots,q$. 
By the definition of reduction $m(S_i)\le f(\alpha){\rm OPT}_{I_i}$.
Then $M_k$ applies $M_{k-1}$ to $S_1,\ldots,S_q$, obtaining a solution $S$.
Then, by the induction hypothesis $m(S)\le f^{k-1}(f(\alpha)){\rm OPT}_I$ and hence $m(S)\le f^{k}(\alpha){\rm OPT}_I$, as required.

The second claim follows easily, since $R_k$ generates $q^k$ instances.
\end{proof}

\section{Maximum Independent Set}

\begin{theorem}
\label{th:indset}
There is a polynomial-time reduction scheme for {\sc Maximum Independent Set} with approximation $a(r,\alpha)=\alpha r$, for any $r \in \mathbb{Q}$, $r>1$. 
\end{theorem}

\begin{proof}
Let $r=\tfrac{k}{l}$, $k,l\in\mathbb{N}$, $k\ge l>0$.
Let $G=(V,E)$ be the input graph.
Reducer partitions $V$ into $k$ parts $V_0,\ldots,V_{k-1}$, of size at most $\lceil|V|/k\rceil$ each.
Then it creates $k$ instances.
Namely, for any $i=0,\ldots,k-1$ it creates $G_i=G[\bigcup_{j=0}^{l-1}V_{(i+j) \bmod k}]$.

Let ${\rm SOL}_0,\ldots,{\rm SOL}_{k-1}$ be solutions (independent sets) for $G_0,\ldots,G_{k-1}$ such that $|{\rm SOL}_i|\ge |{\rm OPT}_{G_i}|/\alpha$.
Merger simply picks the biggest solution.

We claim that for some $i=0,\ldots,k-1$, $|V(G_i)\cap {\rm OPT}|\ge \frac{l}{k}|{\rm OPT}|$.
Indeed, otherwise all of $V(G_i)$ contain less than $l|{\rm OPT}|$ copies of elements of ${\rm OPT}$, with each element of ${\rm OPT}$ appearing in exactly $l$ copies, hence some elements of ${\rm OPT}$ do not belong to any $V(G_i)$, a contradiction. 
Clearly, if $|V(G_i)\cap {\rm OPT}|\ge \frac{l}{k}|{\rm OPT}|$ then $|{\rm OPT}_{G_i}|\ge\frac{l}{k}|{\rm OPT}|={\rm OPT}/r$. 
Hence $|{\rm SOL}_i|\ge |{\rm OPT}|/(\alpha r)$ and the solution returned by merger will be at least that good.
\end{proof}

\section{Coloring}
\label{sec:coloring}

There is a following simple reduction for {\sc Vertex Coloring}.

\begin{theorem}
\label{th:coloring-simple}
There is a polynomial-time reduction scheme for {\sc Vertex Coloring} with approximation $a(r,\alpha)=\alpha r$, for any $r \in \mathbb{N}$, $r>1$.
\end{theorem}

\begin{proof}
Let $G=(V,E)$ be the input graph.
Reducer partitions $V$ into two $r$ sets $V_1,\ldots,V_r$, with at most
$\lceil |V|/r \rceil$ vertices each and returns $r$ instances $G_1,\ldots,G_r$ such that
$G_i=G[V_i]$.

The input for merger is a coloring $c_i:V_r\rightarrow\{1,\ldots,q_r\}$ for each graph $G_i$ such that $q_i\le\alpha\chi(G_i)$.
For any $i$, $G_i\subseteq G$, so $\chi(G_i)\le\chi(G)$ and further $q_i\le\alpha\chi(G)$.
Merger simply colors each $v\in V_i$ with color $\sum_{j=1}^{i-1}q_j + c_i(v)$.
Clearly, it uses at most $\alpha r \chi(G)$ colors, as required.
\end{proof}

A more sophisticated reduction was found by Bj\"orklund and Husfeldt~\cite{bhk:color-j}.
Basically, it removes maximum independent sets (found by an exact algorithm) until the graph is small enough, and the merger just colors the previously removed vertices with new colors (one per each independent set). 
However, taking into account current best time bounds of exact polynomial space algorithms for {\sc Max Independent Set} and {\sc Vertex Coloring}, this reduction makes sense only when the reduction rate $r<4.05$ (roughly).
This is because for larger rates the total time of the resulting approximation is dominated by finding maximum independent sets, so we would get worse approximation guarantee with the same total time.
A natural idea here is to plug our {\sc Max Independent Set} approximation into the algorithm of Bj\"orklund and Husfeldt. 
By modifying their analysis we get the following theorem.

\begin{theorem}
\label{th:coloring}
Assume that there is a $\beta$-approximation algorithm for {\sc Maximum Independent Set} with time complexity $T(n)$.
Then there is an $O^*(T(n))$-time reduction scheme for {\sc Vertex Coloring} with approximation $a(r,\alpha)=\alpha+\beta \ln r$ plus additive error 1, for any $r \in \mathbb{Q}$, $r>1$.
\end{theorem}

\begin{proof}
Let $n=|V(G)|$.
As long as the number of vertices of $G$ exceeds $n/r$ reducer finds an independent set using the $\beta$-approximation algorithm and removes it from $G$.
Let $I_1,\ldots,I_t$ be the independent sets found.
Reducer returns the resulting graph $G'$.

Now let us upperbound $t$.
Let $G_0=G$ and let $G_j$ be the graph obtained from $G_{j-1}$ by removing $I_j$.
Since any subgraph of $G$ is $\chi(G)$-colorable, 
for any $j=1,\ldots,k$, $G_j$ contains an independent set of size at least $|V(G_j)|/\chi(G)$.
It follows that \[|V(G_j)|\le\left(1-\frac{1}{\beta\chi(G)}\right)^j n\le e^{-\frac{j}{\beta\chi(G)}}n.\]
Hence for $j\ge\chi(G)\beta\ln r$ we have $|V(G_j)|\le n/r$ so $t\le\lceil\chi(G)\beta\ln r\rceil$.

Let $c:V(G')\rightarrow \{1,\ldots,q\}$ be a coloring of $G'$, $q\le\alpha\chi(G')$.
Since $G'$ is a subgraph of $G$, $\chi(G')\le\chi(G)$ and hence $q\le\alpha\chi(G)$ colors. 
Merger returns the following coloring of $V(G)$: if $v\in V'$ it has color $c(v)$ and if $v\in I_j$ it has color $q+j$. 
Clearly, it uses at most $\alpha\chi(G)+t\le\alpha\chi(G)+\lceil\chi(G)\beta\ln r\rceil\le (\alpha+\beta\ln r)\chi(G)+1$ colors.
The time complexity of the reduction is $O(nT(n))$.
\end{proof}

\begin{theorem}
\label{cor:color}
Given any $O^*(2^{cn})$-time exact algorithm for {\sc Maximum Independent Set},
for any $\beta\ge 1$, one can construct an  
$O^*(2^{cn/\beta})$-time reduction scheme for {\sc Vertex Coloring} with approximation $a(r,\alpha)=\alpha+\beta \ln r$ plus additive error 1, for any $r \in \mathbb{Q}$, $r>1$.
\end{theorem}
\begin{proof}
The $(\beta,\beta)$-reduction from Theorem~\ref{th:indset} together with the
$O^*(2^{cn})$-time exact algorithm gives $\beta$-approximation for {\sc Maximum Independent Set} in time $O^*(2^{cn/\beta})$.
By Thm.~\ref{th:coloring} we get the claim.
\end{proof}

If we have an $O^*(2^{dn})$-time exact algorithm for {\sc Vertex Coloring},
it makes sense to have a $(a,r)$-reduction of time $O^*(2^{dn/r})$.
After putting $\beta=cr/d$ in Theorem~\ref{cor:color} (and keeping $\beta\ge 1$) we get
$O^*(2^{dn/r})$-time reduction scheme for {\sc Vertex Coloring} with approximation $a(r,\alpha)=\alpha+(cr/d)\ln r$ plus an additive error 1, for any $r \in \mathbb{Q}$, $r\ge d/c$.
With the currently best known value $c=0.288$ and $d=1.167$ (we consider polynomial space here) it gives $a(r,\alpha)=\alpha+0.247r\ln r$ for any $r \in \mathbb{Q}$, $r\ge d/c > 4.05$.

Note also that Theorem~\ref{cor:color} implies that beginning from some value of $r$, the simple $(r,r)$-reduction from Theorem~\ref{th:coloring-simple} outperforms the above more sophisticated one.
Specifically, for the current values of $c$ and $d$ the threshold is $r\ge 58$.

\section{Reduction for Bandwidth}
\label{sec:app:approx-bandwidth}

In this section we describe a $(2,9\alpha)$-reduction for {\sc Bandwidth}.
The following observation will be convenient in our proof.
Observe that any ordering $f:V\rightarrow\{1,\ldots,|V|\}$ corresponds to a sequence $f^{-1}(1),\ldots,f^{-1}(|V|)$, which will be denoted as $s(f)$. Clearly, bandwidth of $f$ corresponds to maximum distance in $s(f)$ between ends of an edge.

\begin{theorem}
There is a polynomial-time $(2,9\alpha)$-reduction for {\sc Bandwidth}.
\end{theorem}

\begin{proof}
Let $G=(V,E)$ be the input graph.
The reducer we are going to describe creates just one instance. 
W.l.o.g. $V$ does not contain isolated vertices for otherwise reducer just removes them
and merger just adds them in the end of the ordering.
We will also assume that ${\rm bw}(G)\ge 2$ -- otherwise reducer replaces $G$ by the empty graph and merger finds the optimal solution in polynomial (linear) time.

Reducer begins with finding a maximum cardinality matching $M$ in $G$.
Next it defines a function $\rho:V\rightarrow V$.
Note that an unmatched vertex has all neighbors matched.
For each unmatched vertex $v$ for each of its (matched) neighbors $w$ we put $\rho(w)=w$.
Consider any $uw\in M$. 
Assume both $\rho(u)=u$ and $\rho(w)=w$. 
Then $u$ has an unmatched neighbor $x$ and $w$ has an unmatched neighbor $y$.
Observe that $x=y$ for otherwise $M\setminus\{uw\}\cup\{xu,wy\}$ is a matching larger than $M$.
In this special case redefine $\rho(u)=w$.
If for some $uw\in M$ both $\rho(u)$ and $\rho(w)$ are unspecified yet we put $\rho(u)=u$ and $\rho(w)=u$ (the choice which endpoint is $u$ and which $w$ is arbitrary).
Finally, if for some $uw\in M$ exactly one value of $\rho$ is specified, say $\rho(u)=u$, we put $\rho(w)=u$. Now $\rho$ is fully defined in subdomain $V(M)$. Note the following claims. 

\noindent
{\bf Claim 1.} For any edge $uw\in M$ either $\rho(u)=u$ and $\rho(w)=u$ or $\rho(u)=w$ and $\rho(w)=w$.

\noindent
{\bf Claim 2.} For any $x\not\in V(M)$, for any its neighbor $w$, $\rho(w)=w$ unless there is a triangle $uwx$ with $uw\in M$.

Now reducer specifies the value of $\rho$ on unmatched vertices, one by one.
Let $x$ be an unmatched vertex with $\rho(x)$ unspecified and let $w$ be any of its neighbors.
If $w$ has another unmatched neighbor $y$ with $\rho(y)$ unspecified we put $\rho(x)=x$ and $\rho(y)=x$. Otherwise we put $\rho(x)=w$.
The way we defined $\rho$ implies the following two claims.

\noindent
{\bf Claim 3.} For at least half of the vertices $\rho(v)\neq v$.

\noindent
{\bf Claim 4.} For any $v\in V$, $|\rho^{-1}(v)|\le 3$.

Finally, reducer simply identifies each pair of vertices $u$, $w$ (i.e.\ adds edges $wx$ such that $x\in N(u)\setminus N[w]$ and removes $u$) such that $\rho(u)=w$.
Denote the resulting graph $G'=(V',E')$. 
Let $V'\subset V$ in the sense that $V'=\{v\in V\ |\ \rho(v)=v\}$.
By Claim 3 the rate of the reduction is at least 2.

Before we describe the merger let us bound ${\rm bw}(G')$.
Let $f:V\rightarrow\{1,\ldots, |V|\}$ be the ordering of $V(G)$ with bandwidth ${\rm bw}(G)$.

\noindent
{\bf Claim 5.} ${\rm bw}(G')\le 3{\rm bw}(G)-1$.

\noindent
{\em Proof of Claim 5.} 
Let $g:V'\rightarrow\{1,\ldots,|V'|\}$  be the ordering of $V'$ which arranges vertices of $V'$ in the same order as $f$ does, i.e.\ $s(g)$ which is obtained from $s(f)$ by removing the vertices outside $V'$.
We will show that $g$ has bandwidth at most $3{\rm bw}(G)-1$.
Consider any $u'v'\in E'$. 
Then for some $u, v\in V$, $\rho(u)=u'$, $\rho(v)=v'$ and $uv\in E$.

If both $f(u)$ and $f(v)$ are outside the interval $(\min\{f(\rho(u)),f(\rho(v))\},\max\{f(\rho(u)),f(\rho(v))\})$, but at opposite sides of it, then $|f(\rho(u))-f(\rho(v))|\le {\rm bw}(G)$.
If both $f(u)$ and $f(v)$ are outside this interval, but at the same side of it, say at $u$'s side, then since $|f(v)-f(\rho(v))|\le 2b$, we have $|f(\rho(u))-f(\rho(v))|\le 2{\rm bw}(G)$.
In any of these two cases, $|f(\rho(u))-f(\rho(v))|\le 2{\rm bw}(G)\le 3{\rm bw}(G)-2$. It follows that $\rho(u)$ and $\rho(v)$ are at distance at most $3{\rm bw}(G)-2$ in $s(g)$, as required.

Now assume one of $f(u)$ and $f(v)$, say $f(u)$, is in the interval $(\min\{f(\rho(u)),f(\rho(v))\},\max\{f(\rho(u)),f(\rho(v))\})$, so in particular $u\not\in V'$. 
Then $\rho(u)$ and $\rho(v)$ are in $s(g)$ at distance smaller by at least 1 from their distance in $s(f)$. Hence it suffices to show that $|f(\rho(u))-f(\rho(v))|\le 3{\rm bw}(G)$.

First note that if $\rho(u),\rho(v)\not\in V(M)$ then by Claim 1 $u,v\not\in V(M)$ and then
$M\cup\{uv\}$ is a matching larger than $M$, a contradiction.
Hence w.l.o.g.\ we can assume that $\rho(u)\in V(M)$.
Then $\rho(u)=u$ or $\rho(u)$ is a neighbor of $u$.

Now assume $\rho(v)\in V(M)$. 
Then also $\rho(v)=v$ or $\rho(v)$ is a neighbor of $v$.
It follows that
$|f(\rho(u))-f(\rho(v))|\le |f(\rho(u))-f(u)|+|f(u)-f(v)|+|f(v)-f(\rho(v))|\le 3{\rm bw}(G)$.

Finally, let $\rho(v)\not\in V(M)$. 
Then $v$ is at distance at most 2 from $\rho(v)$ and hence 
$|f(\rho(v))-f(v)|\le 2{\rm bw}(G)$.
By Claim 2 either $\rho(u)=u$ or $v\rho(u)\in E$ so in any case $v\rho(u)\in E$ which implies
$|f(v)-f(\rho(u))|\le {\rm bw}(G)$.
Together we get $|f(\rho(u))-f(\rho(v))|\le |f(\rho(v))-f(v)|+|f(v)-f(\rho(u))|\le 3{\rm bw}(G)$.
It finishes the proof of Claim 5.

Now we describe the merger.
Let $f':V'\rightarrow \{1,\ldots,|V'|\}$ be the ordering of vertices of $V'$ with bandwidth at most $\alpha {\rm bw}(G')$ for some $\alpha\ge 1$.
By Claim 5, bandwidth of $f$ is at most $\alpha (3{\rm bw}(G)-1)\le 3\alpha{\rm bw}(G)-1$.

Merger returns the ordering $f$ such that $s(f)$ is obtained from $s(f')$ by adding vertices of $\rho^{-1}(v)\setminus\{v\}$ (there are at most 2 of them by Claim 4) right after $v$.
Clearly $s(f)$ is a permutation of $V$.
Now consider any edge $uv\in E$.
There are at most $3\alpha{\rm bw}(G)-2$ vertices between $\rho(u)$ and $\rho(v)$ in $s(f')$.
It follows that there are at most $3(3\alpha{\rm bw}(G)-2)=9\alpha{\rm bw}(G)-6$ vertices between $\rho(u)$ and $\rho(v)$ in $s(f)$.
In other words, the distance between $\rho(u)$ and $\rho(v)$ in $s(f)$ is at most $9\alpha{\rm bw}(G)-5$.
As $u$ is at distance at most 2 from $\rho(u)$ in $s(f)$, and the same holds for $v$, it follows that the distance between $u$ and $v$ in $s(f)$ is at most $9\alpha{\rm bw}(G)-1$.
It follows that $f$ has bandwidth at most $9\alpha{\rm bw}(G)-1$.
\end{proof}

The above theorem together with Reduction Composition Lemma implies:

\begin{corollary}
There is a polynomial-time reduction scheme for {\sc Bandwidth} with approximation $a(r,\alpha)=\alpha 9^k$, for any $r=2^k$, $k\in \mathbb{N}$. 
\end{corollary}

For any $k$, the above reduction gives a $9^k$-approximation in time $O^*(10^{n/2^k})$ and polynomial space (using the exact algorithm of Feige and Kilian), or in time $O^*(5^{n/2^k})$ and $O^*(2^{n/2^k})$ space (using the exact algorithm of Cygan and Pilipczuk).

\section{$O^*(c^n)$-time polynomial space exact algorithms for {\sc Set Cover}}
\label{sec:exact-setcover}
Clearly, to make use of universe-scaling reduction we need a $O^*(c^n)$ exact algorithm, where $c$ is a constant. As far as we know there is no such result published. 
However we can follow the divide-and-conquer approach of Gurevich and Shelah~\cite{gurevich} rediscovered recently Bj\"orklund and Husfeldt~\cite{bjohus:icalp} and we get a $O^*(4^nm^{\log n})$-time algorithm. If $m$ is big we can use another, $O^*(9^n)$-time version of it.

\begin{theorem}
There is a $O^*(\min\{4^nm^{\log n},9^n\})$-time algorithm that finds a mi\-ni\-mum-weight cover of universe of size $n$ by a family $m$ sets.
\end{theorem}

\begin{proof}
The algorithm is as follows. 
For an instance with universe $U$ of size $n$ we recurse on an exponential number of instances, each with universe of size smaller than $n/2$.
Namely, we choose one of $m$ sets $S$ and we divide the remaining elements, i.e. $U\setminus S$ into two parts, each of size at most $n/2$. 
We consider all choices of sets and all such partitions -- there are $O(m2^n)$ of them. 
For each such set $S$ and partition $U_1, U_2$ we find recursively $\mathcal{C}_1$, an optimal cover of $U_1$ and $\mathcal{C}_2$, an optimal cover of $U_2$.
Clearly $\mathcal{C}_1\cup\mathcal{C}_2\cup\{S\}$ forms a cover of $U$.
We choose the best cover out of the $O^*(m2^n)$ covers obtained like this.

Consider an optimal cover ${\rm OPT}$. 
To each element $e$ of $U$ assign a unique set $S_e$ from ${\rm OPT}$ such that $e\in S_e$. 
For each $S \in {\rm OPT}$ let $S^*=\{e \in U: S=S_e\}$. 
Let $\mathcal{P} = \{S^* : S\in {\rm OPT}\}$. 
$\mathcal{P}$ is a partition of $U$.
Clearly, after removing the biggest set $\hat{S}$ from ${\rm OPT}$ we can divide all the sets in $\mathcal{P}$ into two groups, $\mathcal{P}_1$ and $\mathcal{P}_2$, each covering less than $n/2$ elements from $U\setminus\hat{S}$.
It follows that one of the $O^*(m2^n)$ covers found by the algorithm has weight $w({\rm OPT})$, namely the cover obtained for set $\hat{S}$ and partition $(\bigcup\mathcal{P}_1\setminus\hat{S},\bigcup\mathcal{P}_2\setminus\hat{S})$.

It is clear that the above algorithm works in time $O^*(4^nm^{\log n})$.
Similarly we can also get a $O^*(9^n)$ bound -- instead of at most $m2^n$ instances we
recurse on at most $2\cdot3^n$ instances: we consider all partitions of $U$ into three sets $A$, $B$, $C$. Let $|A|\ge|B|\ge|C|$. If $|A| \le n/2$ we recurse on $A$, $B$ and $C$ and otherwise we check whether $A\in\calS$ and if so, we recurse on $B$ and $C$.
\end{proof}

\section{Semi-Metric TSP}
\label{sec:atsp}

{\sc Semi-Metric TSP} is a variant of the classical {\sc Traveling Salesman Problem}.
Here we are also given $n$ vertices and an edge weight function $w:V^2\rightarrow\mathbb{R}$, however now the function $w$ does not need to be symmetric, i.e.\ for any $x,y$ we may have $w(x,y)\ne w(y,x)$. 
Thus the instance can be viewed as a directed graph, with two oppositely oriented edges joining every pair of vertices.
In this variant we assume that $w$ satisfies triangle inequality.
The goal is to find the lightest (directed) Hamiltonian cycle.

In contrast to other problems considered in this paper it is not known whether {\sc Semi-Metric TSP} is in APX. 
The first approximation algorithm for this problem appeared in the work of Frieze, Galbiati and Maffioli~\cite{frieze:logn} and it has approximation ratio $\log_2 n$. 
Currently the best result is a $\frac{2}{3}\log n$-approximation due to Feige and Singh~\cite{feige:atsp}.

The best known exact algorithms are the $O^*(2^n)$-time exponential space classical algorithm by Held and Karp~\cite{held-karp} and a $O^*(4^nn^{\log n})$-time polynomial space algorithm by Bj\"orklund and Husfeldt~\cite{bjohus:icalp}.

The idea of our reduction is very simple -- similarly as in Section~\ref{sec:greedy} we run a polynomial-time approximation, namely the algorithm by Frieze et al.\ and stop it the middle.
%One can do the same also with the algorithm of Feige and Singh but it 
Let us recall the algorithm of Frieze et al.
It begins with finding a lightest cycle cover $C_0$ in $G$ (this can be done in polynomial time by finding a minimum weight matching in a corresponding bipartite graph).
Note that $w(C_0)\le w({\rm OPT}_G)$.
If the cycle cover consists of just one cycle we are done.
Otherwise the algorithm selects one vertex from each cycle. 
Let $G_1$ be the subgraph of graph $G$ induced by these vertices.
Then we find a lightest cycle cover in $G_1$.
Note that $w({\rm OPT}_{G_1}) \le w({\rm OPT}_{G})$, which follows by the triangle inequality.
Then again $w(C_1)\le w({\rm OPT}_G)$.
Again, if $C_1$ has just one cycle we finish, otherwise we choose a vertex from each cycle, build $G_2$ and so on.
As the cycles in cycle covers have lengths at least 2, we finish after finding at most $\log n$ cycle covers. 
Finally we consider the union of all cycle covers $U=\bigcup C_i$. 
Clearly, $U$ is Eulerian and we can find an Eulerian cycle $E$ in $U$.
Then we can transform it to a Hamiltonian cycle $H$ by following $E$, but replacing paths with already visited vertices by single edges. 
We see that $w(H)\le w(E)$ by triangle inequality and hence $w(H)\le \log_2 n \cdot {\rm OPT}$.

Now, assume $r=2^k$. 
If we stop after creating just $k$ cycle covers, we are left with graph $G_k$ with $|V(G_k)|\le |V(G)|/r$, while the cycle covers have weight at most $k\cdot{\rm OPT}$. 
If we have an $\alpha$-approximate TSP tour $T$ in $G_k$, we can add it to the cycle covers and proceed as in the original algorithm. 
Clearly we get a Hamiltonian cycle of weight at most $(\alpha+k){\rm OPT}$.

\begin{corollary}
There is a polynomial-time reduction scheme for {\sc Semi-Metric TSP} with approximation $a(r,\alpha)=\alpha + \log_2 r$, for any $r=2^k$, $k\in \mathbb{N}$. 
\end{corollary}

\end{document}